\documentclass{elsarticle}

\usepackage{lineno,hyperref}
\usepackage{hyperref}

\journal{Nuclear Instruments and Methods in Physics Research A - }

\begin{document}

\begin{frontmatter}

\title{Pre-Production and Quality Assurance of the Mu2e Calorimeter Silicon Photomultipliers}


\author[infn_lnf]{M.~Cordelli}
\author[infn_pisa]{F.~Cervelli}
\author[infn_lnf]{E.~Diociaiuti}
\author[infn_pisa]{S.~Donati}
\author[infn_lnf]{R.~Donghia}
\author[infn_pisa]{S.~Di~Falco}
\author[german]{A.~Ferrari}
\author[infn_lnf]{S.~Giovannella}
\author[infn_lnf]{F.~Happacher}
\author[infn_lnf]{M.~Martini}
\author[infn_pisa]{L.~Morescalchi \corref{mycorrespondingauthor}}
\ead{luca.morescalchi@pi.infn.it}
\author[infn_lnf]{S.~Miscetti}
\author[german]{S.~Muller}
\author[infn_pisa]{E.~Pedreschi}
\author[infn_pisa]{G.~Pezzullo}
\author[infn_lnf]{I.~Sarra}
\author[infn_pisa]{F.~Spinella}

\address[infn_pisa]{INFN Sezione di Pisa, Pisa, Italy}
\address[infn_lnf]{Laboratori Nazionali di Frascati dell'INFN, Frascati, Italy }
\address[german]{HZDR, Helmholtz-Zentrum Dresden-Rossendorf, Germany}
\cortext[mycorrespondingauthor]{Corresponding author}

\begin{abstract}
The Mu2e electromagnetic calorimeter has to provide precise information on energy, time and position for $\sim$100 MeV electrons. It is composed of 1348 un-doped CsI crystals, each coupled to two large area Silicon Photomultipliers (SiPMs). A modular and custom SiPM layout consisting of a 3$\times$2 array of 6$\times$6 mm$^2$ UV-extended monolithic SiPMs has been developed to fulfill the Mu2e calorimeter requirements and a pre-production of 150 prototypes has been procured by three international firms (Hamamatsu, SensL and Advansid). A detailed quality assurance process has been carried out on this first batch of photosensors: the breakdown voltage, the gain, the quenching time, the dark current and the Photon Detection Efficiency (PDE) have been determined for each monolithic cell of each SiPMs array. One sample for each vendor has been exposed to a neutron fluency up to $\sim$8.5~$\times$~10$^{11}$ 1 MeV (Si) eq. n/cm$^{2}$ and a linear increase of the dark current up to tens of mA has been observed. Others 5 samples for each vendor have undergone an accelerated aging in order to verify a Mean Time To Failure (MTTF) higher than $\sim$10$^{6}$ hours. 
\end{abstract}


\end{frontmatter}


\section{Introduction}

The Mu2e Experiment \cite{mu2etdr} will search for the Charged Lepton Flavour Violation (CLFV) coherent conversion of muon into electron in the field of an aluminum nucleus with an unprecedented accuracy, allowing to indirectly probe energy scales up to thousands TeV. One of the most important pieces of the Mu2e detector is the electromagnetic calorimeter~\cite{Atanov:2017dwk}: it consists of 1348 un-doped CsI crystals each coupled to two large area Silicon Photomultipliers (SiPMs) and arranged in two disks. The calorimeter is hosted in a cryostat inside a superconductive solenoid and has to operate in a 10$^{-4}$ Torr vacuum and a 1 T magnetic field. Moreover, it also has to stand the high radiation fluxes coming from the muons stopping target: in the hottest regions, i.e. the inner crystals of the front disk, the radiation level will reach about 10 krad/year and a neutron fluence of $\sim$2$\times$10$^{11}$ n/cm$^2$/yr. \smallskip

The SiPMs must have a good quantum efficiency at 315 nm for optimal coupling with the CsI scintillation emission. Since the detector will be accessible only once a year, the photosensors must have a good reliability so as not to compromise the calorimeter performances with any failure. \smallskip

To fulfill the calorimeter requirements, a custom SiPM layout consisting of a 3$\times$2 array of new generation 6$\times$6 mm$^2$ UV-extended monolithic SiPMs has been designed. The readout is organized as the parallel connection of two series of three monolithic cells. The connection in series of three SiPMs allows to have a large active area with a reduced equivalent capacitance. In this way it is possible to increase the light collection and also to obtain narrowed signals useful to handle the pileup. On the other hand, the bias voltage of the series triples with respect to the one of a single SiPM. 

\section{Quality Assurance of SiPMs Pre-Production}

The Quality Assurance (QA) process for the photosensors selection is requested to detect any device with operative performances under the standards. The QA will characterize the photosensors at the level of the single cell before the assembling in the calorimeter. \smallskip

QA criteria have been fixed starting by the request to have a good uniformity between the cells of the same sensor and to have a light collection of at least 20 photo-electrons/MeV as suggested by simulation. Defining the operational voltage V$_{\rm op}$ as 3 V over the breakdown voltage V$_{\rm br}$, the requirements at a temperature of 20$^{\circ}$ C are: \smallskip

\begin{itemize}
  \item a spread in the breakdown voltage V$_{\rm br}$ between the sensor cells $<$ 0.5\%;
  \item a spread in the dark current at V$_{\rm op}$ between sensor cells $<$ 15\%;
  \item a gain at V$_{\rm op}$ (measured in 150 ns gate) $>$ 10$^6$ for each cell;
  \item a PDE at V$_{\rm op}$ $>$ 20\% for 315 nm;\smallskip
\end{itemize}

In order to perform the final photosensor choice, 150 custom prototypes has been purchased from three international vendors: Hamamatsu and SensL, with a pixel size of 50$\mu$m, and AdvanSid, with a pixel size of 30$\mu$m. This first batches has been tested according to the QA procedure described below.\smallskip

In view of the large number of measurement to perform, a custom semi-automatized system controlled by computer has been developed. The station keeps the temperature of the Sensor Under Test (SUT) stable at 20$^{\circ}$ C. The temperature is continuously monitored by a digital thermometer system with an accuracy of 0.3$^{\circ}$ C. \smallskip

\begin{figure}[htbp]
\centering 
\includegraphics[width=0.95\textwidth]{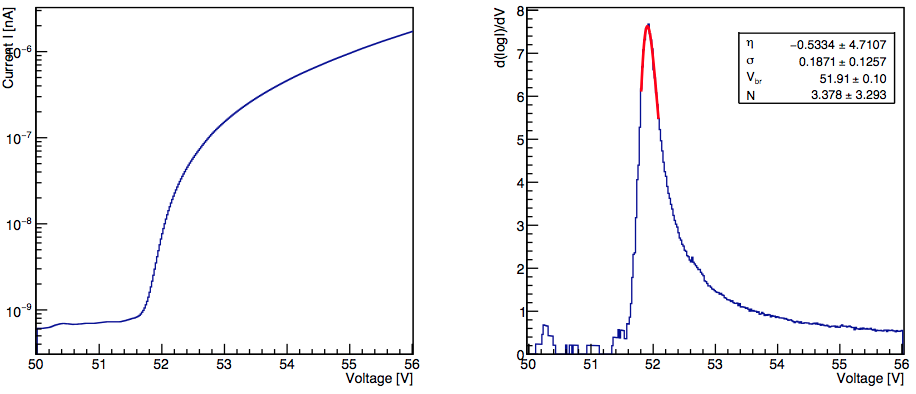}
\caption{\label{fig:ivscan} \textbf{Left} - Example of an I-V scan for a cell. \textbf{Right} - Logarithmic derivative of the I-V curve used to evaluate V$_{\rm br}$.\smallskip}
\end{figure}

First, the I-V dark curve is measured with a 50 mV step resolution. The V$_{\rm br}$ is then obtained by constructing the dlog(I)/dV curve and by fitting the peak position \cite{xu2014study}. An example of this procedure is shown in Figure \ref{fig:ivscan}. The dark current at V$_{\rm op}$ is then easily extracted from the I-V dark curve. \smallskip

To evaluate the gain, the SUT is illuminated with an UV LED driven by 20 ns pulses at 100 kHz frequency. The pulse amplitude is tuned to let only few photons hitting the sensor. The charge is reconstructed by integrating the first 150 ns of signal. The gain is then obtained by taking the difference between the position of the first and the second peaks in the charge distribution, corresponding respectively to 0 and 1 photons hitting the sensor. \smallskip

\begin{figure}[htbp]
\centering 
\includegraphics[width=0.43\textwidth]{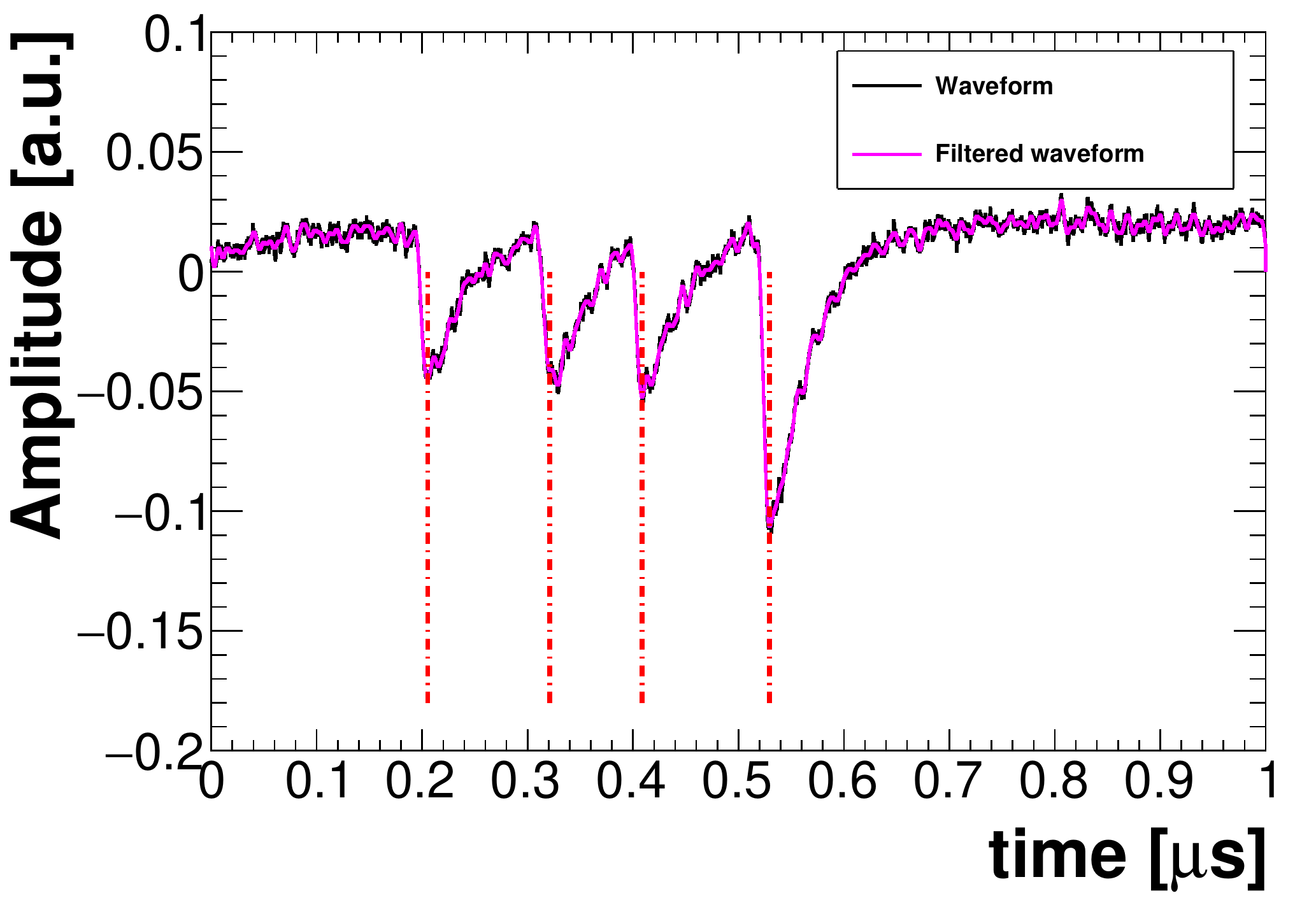} \quad
\includegraphics[width=0.53\textwidth]{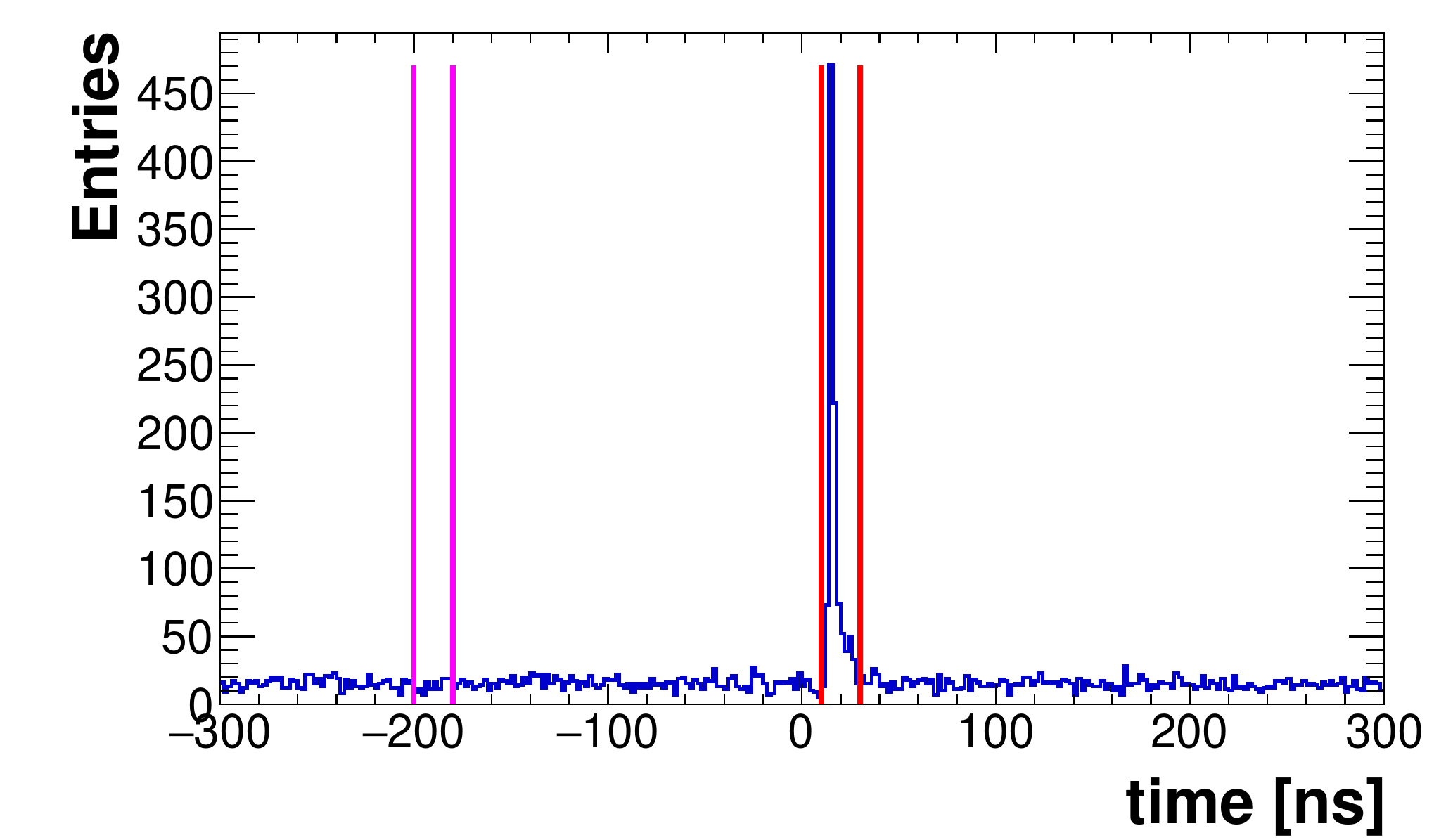}
\caption{\label{fig:pdemeth} \textbf{Left} - In black an example of acquired waveform. In magenta is reported the same waveform obtained by cutting the Fourier components larger than 150 MHz. Red lines indicate the reconstructed time of detected peaks \textbf{Right} - Time of the detected peaks with respect to the LED pulse: the peak between 10 and 30 ns corresponds to the LED photons events, the 
at background out of time is due to the dark pulses. \smallskip} 
\end{figure}

The PDE is determined using a counting method \cite{phdtheis} that directly analyzes all the waveforms triggered in time with a few-photons LED pulse. First, a peak search algorithm is applied to determine the time of each dark or LED signal as in Figure \ref{fig:pdemeth} Left. In the distribution  
of the peaks times (see Figure \ref{fig:pdemeth} Right), two time gates 20 ns large are selected: the first (magenta lines) is used to evaluate the number of dark pulses in the time interval (N$_{D}$), while the second (red lines) is used to estimate the number of events in time with the LED with at least one photoelectron (N$_{n>1}$). Using N$_{D}$, N$_{n>1}$ and the total number of recorded events N$_{T}$, the mean number of detected photoelectrons is obtained as follows:

\begin{equation}
n_{pe} = - ln( 1 - \frac{N_{n\geq 1}}{N_{T}} ) + ln( 1 - \frac{N_D}{N_{T}} )
\end{equation} \smallskip

This method provided the relative PDE with respect to a 3x3 UV-extended SiPM from Hamamtsu used as reference sensor, biased at the operative voltage and previously calibrated with a photo-diode. The stability of the light has been continuously monitored using another sensor posed close to the LED. 

The results of the characterization of the devices from the three vendors are listed in Table \ref{tab:sipm_char}: all of them satisfied the Mu2e technical requirements. \smallskip


\begin{table}[h!]
\centering
\begin{tabular}{|c|c|c|c|}
\hline
 & Hamamatsu & SensL & AdvanSid\\ \hline
V$_{\rm br}$ & (51.85 $\pm$ 0.11) V & (24.87 $\pm$ 0.06) V & (27.20 $\pm$ 0.04) V \\ \hline
RMS(V$_{\rm br})$ & (0.070 $\pm$ 0.005)$\%$ & (0.13 $\pm$ 0.01)$\%$ & (0.11 $\pm$ 0.01)$\%$ \\ \hline
$I_{\rm dark}$ & (0.77 $\pm$ 0.13) $\mu$A & (1.22 $\pm$ 0.28) $\mu$A & (1.07 $\pm$ 0.08) $\mu$A \\ \hline
RMS(I$_{\rm dark})$ & (6.4 $\pm$ 0.5)$\%$ & (8.1 $\pm$ 0.8)$\%$ & (4.7 $\pm$ 0.4)$\%$ \\ \hline
Gain in 150 ns & (2.40 $\pm$ 0.01)$\cdot$10$^6$ & (1.92 $\pm$ 0.01)$\cdot$10$^6$ & (1.10 $\pm$ 0.05)$\cdot$10$^6$  \\ \hline
RMS(Gain) & (1.7 $\pm$ 0.2)$\%$ & (4.3 $\pm$ 0.5)$\%$ & (8.5 $\pm$ 0.7)$\%$ \\ \hline
PDE @ 315 nm & (28.0 $\pm$ 1.2)$\%$  & (32.4 $\pm$ 1.4)$\%$  & (21.3 $\pm$ 0.9)$\%$  \\ \hline

\end{tabular}
\caption{Results of the Mu2e custom SiPMs pre-production characterization for the three vendors. The RMS values are referred to the spread of each parameter between the sensor cells.}
\label{tab:sipm_char}
\end{table}

\section{Increase of Dark Current due to Radiation Damage}

\begin{figure}[htbp]
\centering 
\includegraphics[width=0.49\textwidth]{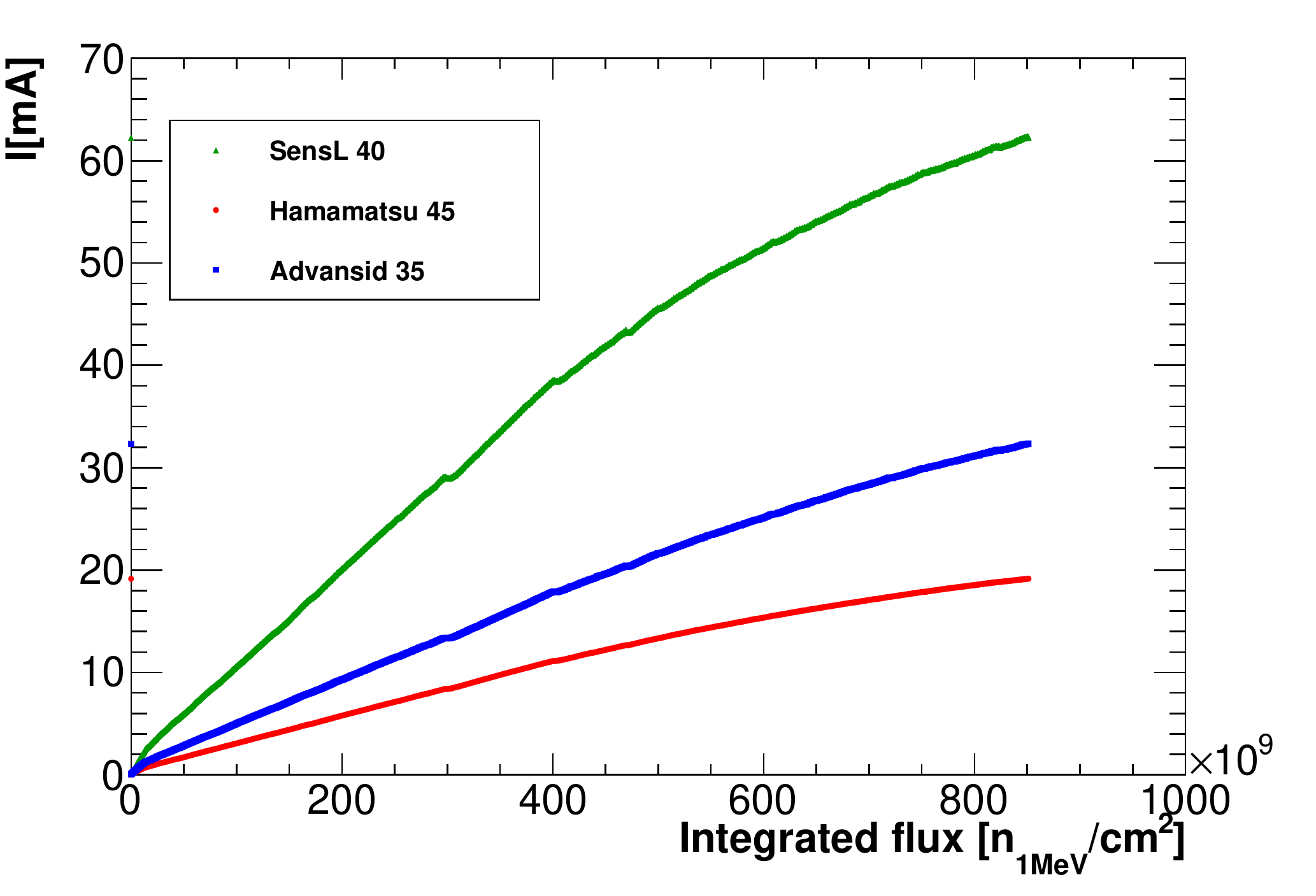}
\includegraphics[width=0.5\textwidth]{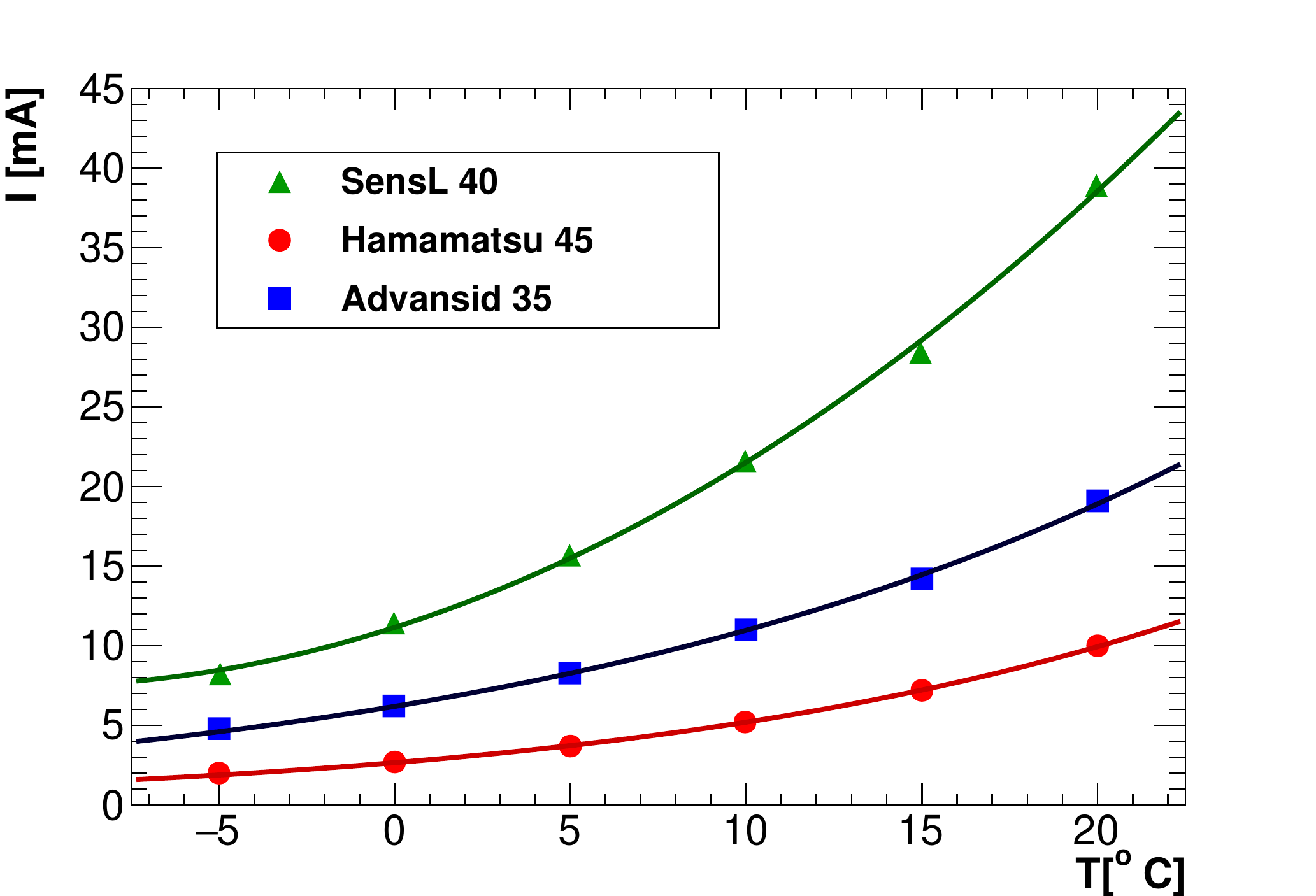}
\caption{\label{fig:irradiation} \textbf{Left} - Dark current of a sensor cell as a function of the neutron fluence. Deviation from linear dependence at higher fluence are due to voltage drop on the cable. \textbf{Right} - Dark current of a sensor cell as a function of the temperature after $\sim$8.5~$\times$~10$^{11}$ 1 MeV (Si) eq. n/cm$^{2}$ irradiation.}
\end{figure}

Radiation damage can create defects in silicon detectors, which mainly increase the dark current \cite{leroy2007particle}. Simulation studies estimated that, in the highest irradiated regions, each photosensor will absorb a dose of 20 krad and will be exposed to a neutron fluence of $\sim$ 8~$\times$~10~$^{11}$ 1 MeV (Si) eq. n/cm$^{2}$ in three years of running, with a safety factor of three to take into account  uncertainties in the Montecarlo simulation. Since for these fluxes the damage dealt by ionizing particles is negligible with respect to the displacement damage due to neutron interactions \cite{baccaro2017irradiation}, the photosensors have been tested with the neutron generated by the EPOS facility of HZDR in Dresden. This facility can provide a clean neutron flux centered at 1 MeV with  negligible photon contamination. One device/vendor has been exposed to a fluence up to $\sim$8.5~$\times$~10$^{11}$ 1 MeV (Si) eq. n/cm$^{2}$ over $\sim$29 hours. The SUTs have been cooled at 20$^{\circ}$ C and biased at their operational voltage, determined as explained above. The dark current has been continuously measured during the irradiation. A linear increase of the dark current as a function of the fluence with a different slope between vendors has been observed (see Figure \ref{fig:irradiation} Left). 

For the three vendors, the dependence of the dark current at V$_{\rm op}$ from the temperature after the irradiation is shown in Figure \ref{fig:irradiation} Right. To limit the dark current, the SiPMs will be operated at a temperature of 0$^{\circ}$~C.

\section{Mean Time To Failure}

Each of the two sensors coupled to the same crystal can independently satisfy the request on the light collection. In this way, to lose a calorimeter channel both the sensors have to fail. The Mean Time To Failure (MTTF) needed to maintain a fully performing calorimeter along the planned three years of running is of the order of $\sim$10$^6$ hours/component. In order to obtain an MTTF experimental estimation for the Mu2e custom SiPMs, 5 sensors/vendor have been subjected to accelerated aging. These sensors have been stressed by operating them at V$_{\rm op}$ inside a light tight box kept at a temperature of 50$^{\circ}$~C. According to the Arrhenius Equation \cite{vigrass2010}, this temperature corresponds to an acceleration factor of $\sim$100. During the 2500 hours of test the sensors were continuously monitored by controlling their response to a pulsed LED every 2 minutes and by registering the behavior of the dark current in time. No dead sensors have been observed for all the three vendors, confirming an MTTF value greater than 0.65~$\times$~10$^6$ hours. 

\section{Conclusions}

A first batch of 150 custom photosensor prototypes from three different vendors has been fully characterized and tested both for radiation hardness and reliability. These results helped to define the QA procedure to test the photosensors production: this QA process will involve more than 3000 devices, for a total of more than 18000 monolithic cells. After completing QA, the photosensors will be assembled together with the cystals in the calorimeter disks.    

\section*{Acknowledgments}
 This work was supported by the EU Horizon2020 Research and Innovation Programme under the Marie Sklodowska-Curie Grant Agreement No. 690835.

\section*{References}

\bibliography{mybibfile}

\end{document}